\documentclass[prl,aps,twocolumn,footinbib,superscriptaddress]{revtex4-1}
\usepackage{amsmath}
\usepackage{amssymb}
\usepackage{graphicx}
\usepackage{dcolumn}
\usepackage{enumerate}
\usepackage{bm}
\usepackage{xcolor}
\usepackage{comment}
\usepackage{subfigure}
\usepackage{breqn}
\usepackage{hyperref}
\hypersetup{colorlinks=true, citecolor=blue, urlcolor=blue, linkcolor=blue}

\makeatletter
\let\cat@comma@active\@empty
\makeatother

\begin{document}
\title{Simulation of XXZ Spin Models using Sideband Transitions in Trapped Bosonic Gases}
\author{Anjun Chu}
\email{anjun.chu@colorado.edu}
\affiliation{JILA, NIST and Department of Physics, University of Colorado, Boulder, CO 80309, USA}
\affiliation{Center for Theory of Quantum Matter, University of Colorado, Boulder, CO 80309, USA}
\author{Johannes Will}
\affiliation{Institut f\"{u}r Quantenoptik, Leibniz Universit\"{a}t Hannover, Welfengarten 1, D-30167 Hannover, Germany}
\author{Jan Arlt}
\affiliation{Institut for Fysik og Astronomi, Aarhus Universitet, Ny Munkegade 120, DK-8000 Aarhus C, Denmark}
\author{Carsten Klempt}
\affiliation{Institut f\"{u}r Quantenoptik, Leibniz Universit\"{a}t Hannover, Welfengarten 1, D-30167 Hannover, Germany}
\author{Ana Maria Rey}
\affiliation{JILA, NIST and Department of Physics, University of Colorado, Boulder, CO 80309, USA}
\affiliation{Center for Theory of Quantum Matter, University of Colorado, Boulder, CO 80309, USA}
\date{\today}

\begin{abstract}
We theoretically propose and experimentally demonstrate the use of motional sidebands in a trapped ensemble of $^{87}$Rb atoms to engineer tunable long-range XXZ spin models. We benchmark our simulator by probing a ferromagnetic to paramagnetic dynamical phase transition in the Lipkin-Meshkov-Glick (LMG) model, a collective XXZ model plus additional transverse and longitudinal fields, via Rabi spectroscopy. 
We experimentally reconstruct the boundary between the dynamical phases, which is in good agreement with mean-field theoretical predictions. 
Our work introduces new possibilities in quantum simulation of anisotropic spin-spin interactions and quantum metrology enhanced by many-body entanglement.
\end{abstract}

\maketitle

Quantum simulation of iconic models of quantum magnetism in highly controllable atomic systems is emerging as a promising way to gain new insights into fundamental many-body phenomena in condensed matter physics \cite{auerbach1994}, and as a pathway to shed light onto exciting new phenomena in non-equilibrium many-body spin arrays \cite{zhang2017,bernien2017,jurcevic2017,smale2019,yang2019,muniz2020}. 
In recent years, rapid progress in the simulation of quantum spin models has been made by taking advantage of the diversity of interactions in ultracold quantum systems, including contact interactions in the motional ground state of ultracold atomic gases \cite{yang2019,Gross2017}, dipolar interactions in polar molecules \cite{Bohn2017}, magnetic atoms \cite{burdick2016,baier2016,lepoutre2019} and Rydberg atoms \cite{Adams2019}, as well as photon/phonon-mediated long-range interactions in trapped ions \cite{Bruzewicz2019} and cavity QED systems \cite{mottl2012,norcia2018,davis2019,muniz2020,Yudan2019}.
  
One promising avenue in this direction is the fact that non-degenerate thermal gases interacting via purely contact interactions, can emulate spin models by mapping the single-particle energy eigenstates onto a lattice in mode space \cite{smale2019,rey2009,rey2014,koller2016}. 
This mapping has been shown to be a powerful way to emulate long-range interacting spin models featuring large many-body energy gaps that have enabled significant enhancement of coherence time \cite{deutsch2010,buning2011,solaro2016}.
Nevertheless, the tunability of the spin model parameters has so far been mainly accomplished by the use of Feshbach resonances, and the atom loss associated with the latter imposes a trade-off between tunability and coherence time \cite{widera2008,smale2019}.
 
In this Letter, we theoretically propose and experimentally demonstrate the use of motional sidebands in a thermal trapped gas of $^{87}$Rb atoms to engineer long-range XXZ spin models with tunable spin couplings. 
We benchmark our simulator by probing a dynamical phase transition (DPT) between ferromagnetic and paramagnetic phases in the collective XXZ model plus additional transverse and longitudinal fields (also known as the  Lipkin-Meshkov-Glick (LMG) model \cite{lipkin1965, muniz2020,Borish2020,zhang2017}) via Rabi spectroscopy. 
We experimentally reconstruct the boundary of the dynamical phases by varying atom density and longitudinal field strength and show good agreement with mean-field theoretical predictions. 
At the end we also discuss the further applications of our scheme in entanglement-enhanced metrology \cite{wineland1992,kitagawa1993,ma2011}, as well as generalizations to a wide range of quantum systems.

\begin{figure}[t]
    \includegraphics{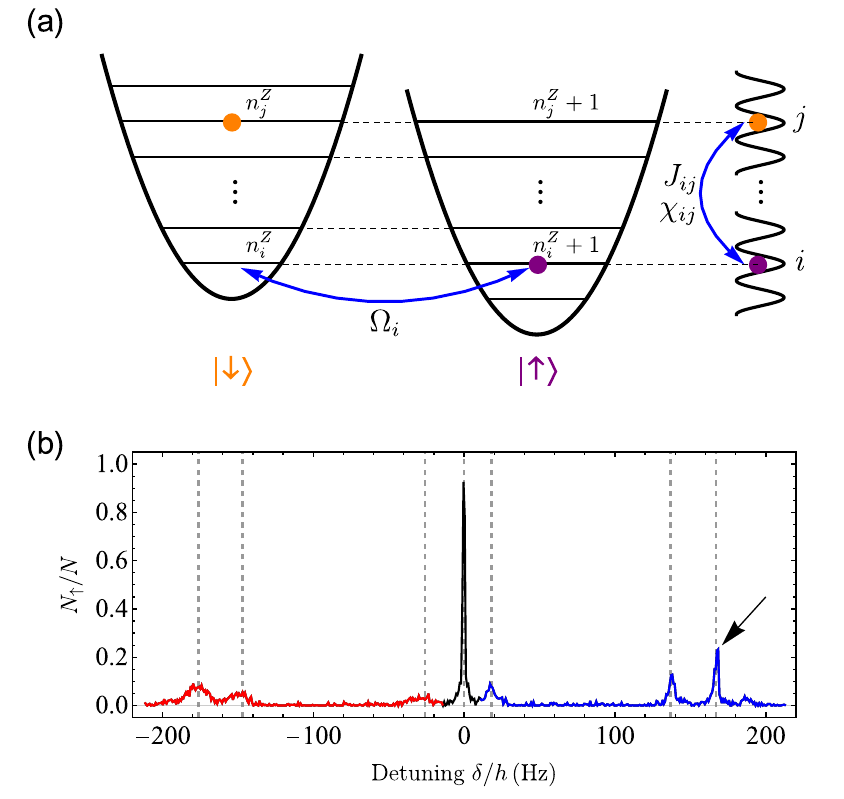}
    \caption{\label{fig1}Simulating XXZ spin models using sideband transitions in a thermal bosonic gas confined in a 3D harmonic trap. (a) Schematic of the effective 3D mode-space lattice for blue $Z$-sideband (only the projection along the $Z$-direction is shown for simplicity). The states $|\Uparrow_i\rangle\equiv|\uparrow;n^X_i,n^Y_i,n^Z_i+1\rangle$ and $|\Downarrow_i\rangle\equiv|\downarrow;n^X_i,n^Y_i,n^Z_i\rangle$, which are the ones coupled by the Raman pulse with Rabi frequency $\Omega_i$, can be regarded as  the two spin states pinned at the $i^{th}$  site of the effective 3D mode-space lattice. Contact interactions in bosonic gases generate long-range XXZ couplings $J_{ij}, \chi_{ij}$ between lattice sites $i$ and $j$ in mode space (see text). (b) Rabi spectrum in the resolved sideband limit for mean atom density $n=2.0\times 10^{12}\mathrm{cm}^{-3}$. The black (blue, red) line represents carrier (blue sideband, red sideband) transitions. Our experiment focuses on the strongest sideband pointed out by the arrow. It is worth mentioning that the suppression of red sideband transitions is related to the anharmonic corrections in optical dipole traps, also observed in Ref.~\cite{allard2016}.}
\end{figure}

\begin{figure}[t]
    \includegraphics{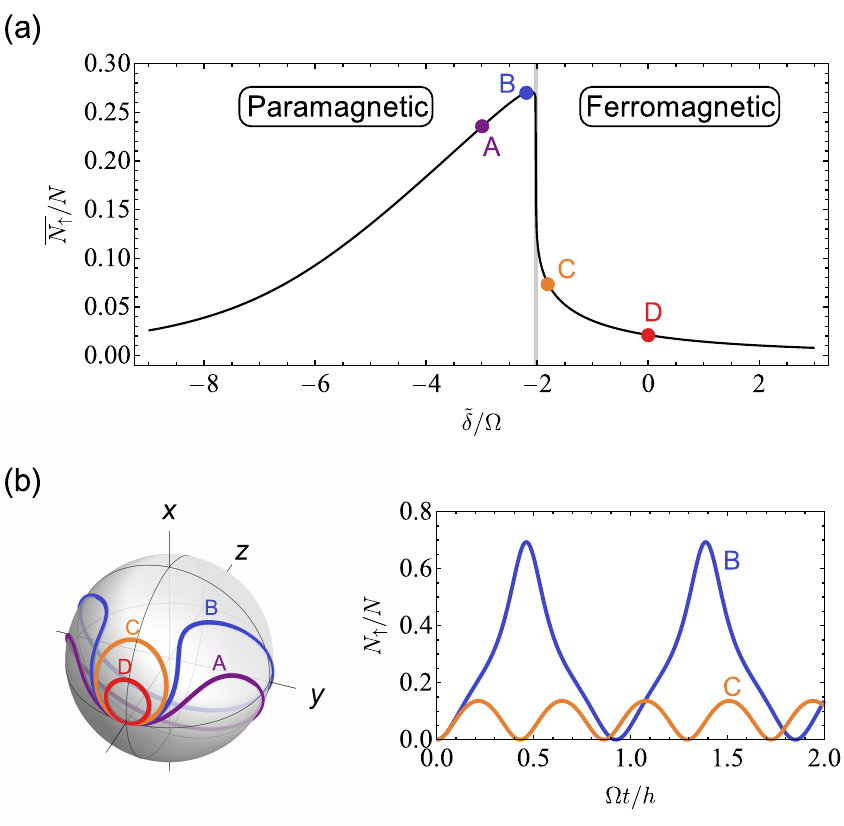}
    \caption{\label{fig2}(a) Dynamical phase transition (DPT) in LMG model with $N\chi/\Omega=5$, indicated by the sharp behavior in long-time average excitation fraction $\overline{N_{\uparrow}}/N$. The critical point is marked by the vertical gray line at $\tilde{\delta}/\Omega=-2.02$, separating the dynamical paramagnetic phase (left) and the dynamical ferromagnetic phase (right). (b) Mean-field dynamics of the LMG model with $\tilde{\delta}/\Omega=-3$(A), $-2.2$(B), $-1.8$(C), $0$(D). The left panel shows the mean-field trajectories on the Bloch sphere, and the right panel presents the mean-field evolution of the excitation fraction for trajectory B and C. The sharp change in dynamics between trajectory B and C also signals the DPT.}
\end{figure}

\begin{figure*}[t]
    \includegraphics{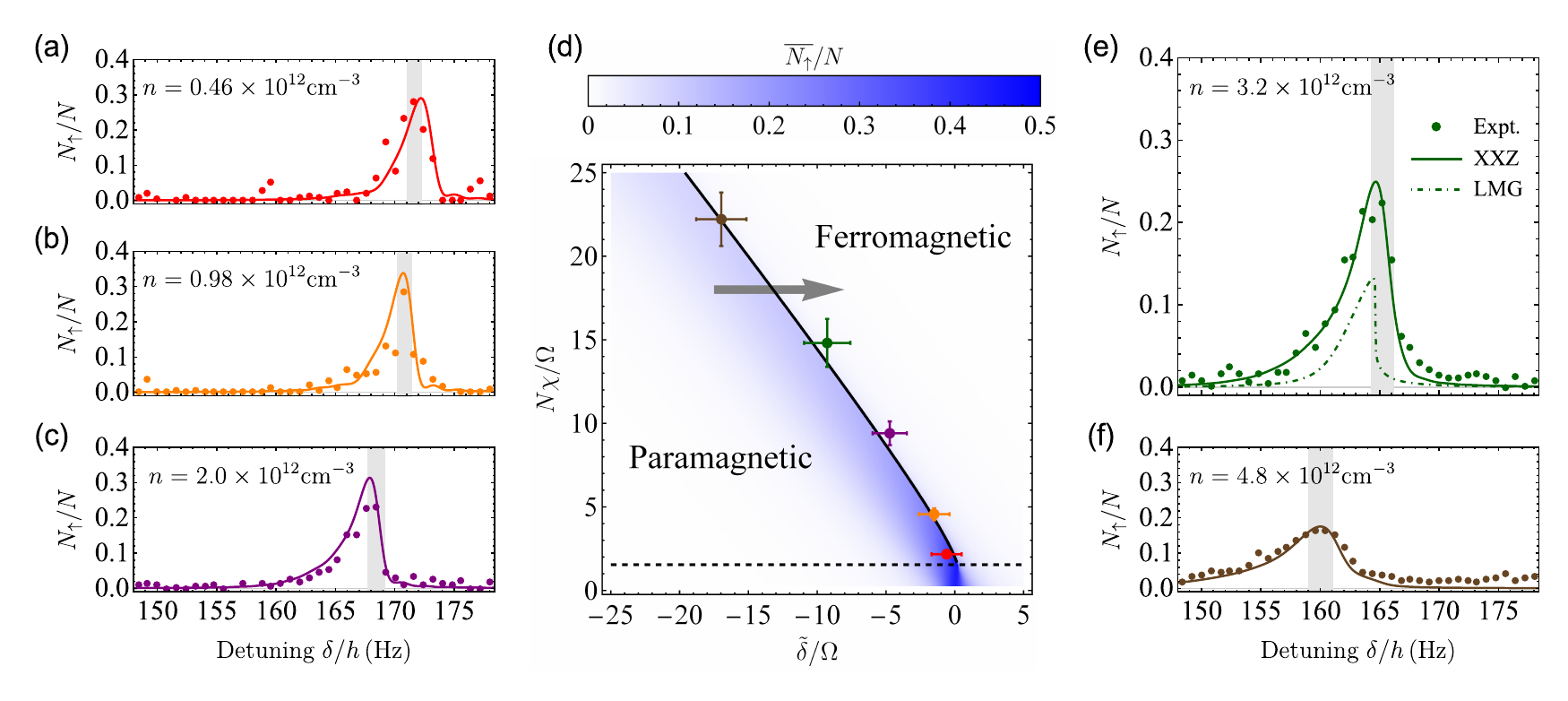}
    \caption{\label{fig3}(a-c, e-f) Dynamical phase transition in the $171$~Hz blue sideband with mean atom density $n=\{0.46,0.98,2.0,3.2,4.8\}\times 10^{12}\mathrm{cm}^{-3}$, indicated by the asymmetric lineshape after evolution time $0.5$s. The shaded areas indicate the critical points, where the uncertainty is set by finite frequency step of detuning as well as fluctuations in atom density and Rabi frequency. The filled circles denote experimental data, the solid lines denote mean-field theoretical predictions by $H_{\mathrm{XXZ}}$, and the green dot-dashed line in (e) denotes the order parameter $\overline{N_{\uparrow}}/N$ predicted by $H_{\mathrm{LMG}}$ (see text). We do not directly add the experimental error bars to the lineshape data in (a-c, e-f) for visual reasons, and the typical statistical uncertainty in each figure is $\Delta N_{\uparrow}/N=\{0.038,0.020,0.013,0.010,0.011\}$ respectively. (d) Phase diagram for ferromagnetic to paramagnetic dynamical phase transition. The black solid line denotes the sharp phase boundary of DPT, the black dashed line separates the smooth crossover regime (below) with DPT regime (above), and the gray arrow illustrates the probing direction on phase diagram. The phase boundary is reconstructed from the critical points in (a-c, e-f) using the same choice of color to label data points.}
\end{figure*}

We consider an ensemble of thermal $^{87}$Rb atoms confined in a 3D harmonic trap and prepared in the magnetically insensitive clock states $|\downarrow\rangle\equiv|F=1, m_F=0\rangle$ and $|\uparrow\rangle\equiv|F=2, m_F=0\rangle$.
The contact interaction in this two-component bosonic gas can be written in the following second quantized form \cite{sorensen2001,micheli2003},
\begin{equation}
    H_{\mathrm{int}}=\sum_{\sigma\sigma'=\uparrow,\downarrow}\frac{U_{\sigma\sigma'}}{2}\int\mathrm{d}^3\mathbf{R}\,\psi^{\dag}_{\sigma}(\mathbf{R})\psi^{\dag}_{\sigma'}(\mathbf{R})\psi_{\sigma'}(\mathbf{R})\psi_{\sigma}(\mathbf{R}),
    \label{eq:contact}
\end{equation}
where $U_{\sigma\sigma'}=4\pi\hbar^2a_{\sigma\sigma'}/m$ is the interaction strength between atoms of spin $\sigma$ and $\sigma'$, parametrized by the $s$-wave scattering lengths, $a_{\uparrow\uparrow}=94.55a_0$, $a_{\uparrow\downarrow}=98.09a_0$, $a_{\downarrow\downarrow}=100.76a_0$ \cite{van2002}. 
The bosonic field operator $\psi_{\sigma}(\mathbf{R})$, is expanded in the eigenmode basis of the 3D harmonic trap, $\psi_{\sigma}(\mathbf{R})=\sum_{\mathbf{n}}a_{\mathbf{n}\sigma}\phi_{\mathbf{n}}(\mathbf{R})$, where $a_{\mathbf{n}\sigma}$ annihilates a boson of spin $\sigma$ in eigenmode $\mathbf{n}=\{n^X,n^Y,n^Z\}$ of the harmonic trap, with corresponding wave function $\phi_{\mathbf{n}}(\mathbf{R})$.

We understand and analyze the many-body dynamics through a mapping of the single-particle eigenstates of the 3D harmonic trap onto a 3D lattice in mode space, as depicted in Fig.~\ref{fig1}(a).
Notice that a blue sideband transition along the $Z$-direction couples the following two states in the harmonic trap, $|\Uparrow_i\rangle\equiv|\uparrow;n^X_i,n^Y_i,n^Z_i+1\rangle$ and $|\Downarrow_i\rangle\equiv|\downarrow;n^X_i,n^Y_i,n^Z_i\rangle$. 
So we can visualize the states  $|\Uparrow_i\rangle$ and $|\Downarrow_i\rangle$ as two spin states localized at site $i$ in an effective 3D mode-space lattice. 
The wave functions associated with $|\Uparrow_i\rangle$ and $|\Downarrow_i\rangle$ states are denoted as $\phi_i^{\Uparrow}(\mathbf{R})$ and $\phi_i^{\Downarrow}(\mathbf{R})$ respectively.
Similar treatments can apply to blue sideband transitions along other directions, carrier transitions as well as red sideband transitions \footnote{See Supplemental Material at [URL will be inserted by publisher] for details of numerical simulations and experimental techniques, includes Ref.~\cite{muniz2020,buning2011,allard2016,ma2011,Lin2009,Reinaudi2007,Maineult2012,Kaplan2002,schachenmayer2015,wineland1992}}.

Since we are interested in the collisionless regime of a trapped atomic ensemble, where the trapping potential is much larger than the interaction strength, we assume that each atom is fixed in the mode-space lattice \cite{smale2019,rey2009,rey2014,koller2016}, and that the only relevant process between two colliding atoms is to either remain in the same internal states or to exchange them. 
Furthermore, we can restrict our discussions to include either empty or singly-occupied lattice sites since the $^{87}$Rb gas temperature is above quantum degeneracy \cite{Note1}.
These approximations map the contact interaction term in the Hamiltonian (see Eq.(\ref{eq:contact})) to a spin-$1/2$ long-range XXZ model in the mode-space lattice:
\begin{equation}
    H_{\mathrm{int}}=\sum_{ij}J_{ij}\mathbf{S}_i\cdot\mathbf{S}_j+\sum_{ij}\chi_{ij}S_i^zS_j^z+\sum_iB_iS_i^z.
    \label{eq:xxz}
\end{equation}
Here, the spin operators can be written in terms of bosonic operators on each lattice site, $\mathbf{S}_i=\sum_{\alpha\beta=\Uparrow,\Downarrow}a^{\dag}_{i\alpha}\bm{\sigma}_{\alpha\beta}a_{i\beta}/2$, where $\bm{\sigma}_{\alpha\beta}$ are Pauli matrices, and $a_{i\beta}$ annihilates a boson of spin $\beta$ on lattice site $i$. 
The XXZ interaction parameters are given by $J_{ij}=V_{ij}^{\mathrm{ex}}U_{\uparrow\downarrow}$, $\chi_{ij}=V_{ij}^{\Uparrow\Uparrow}U_{\uparrow\uparrow}+V_{ij}^{\Downarrow\Downarrow}U_{\downarrow\downarrow}-V_{ij}^{\Uparrow\Downarrow}U_{\uparrow\downarrow}-V_{ij}^{\mathrm{ex}}U_{\uparrow\downarrow}$, and $B_i=\sum_{j\neq i}(V_{ij}^{\Uparrow\Uparrow}U_{\uparrow\uparrow}-V_{ij}^{\Downarrow\Downarrow}U_{\downarrow\downarrow})$, and are set by the overlap integral of the relevant 3D harmonic oscillator wave functions: $V^{\alpha\beta}_{ij}=\int\mathrm{d}^3\mathbf{R}[\phi_i^{\alpha}(\mathbf{R})]^2[\phi_j^{\beta}(\mathbf{R})]^2$, and $V^{\mathrm{ex}}_{ij}=\int\mathrm{d}^3\mathbf{R}\phi_i^{\Uparrow}(\mathbf{R})\phi_i^{\Downarrow}(\mathbf{R})\phi_j^{\Uparrow}(\mathbf{R})\phi_j^{\Downarrow}(\mathbf{R})$.
The tunability of spin-spin couplings depends on these overlap integrals. For carrier transitions we have $\phi_i^{\Uparrow}(\mathbf{R})=\phi_i^{\Downarrow}(\mathbf{R})=\langle \mathbf{R}|n^X_i,n^Y_i,n^Z_i\rangle$, and therefore $V_{ij}^{\alpha\beta}=V_{ij}^{\mathrm{ex}}$, making the XXZ spin model equivalent to the isotropic Heisenberg model ($J_{ij}\gg \chi_{ij}$). For the sideband transitions, the wave functions are not the same for the two spin components (e.g. for the blue $Z$-sideband 
$\phi_i^{\Downarrow}(\mathbf{R})=\langle \mathbf{R}|n^X_i,n^Y_i,n^Z_i\rangle$ and  $\phi_i^{\Uparrow}(\mathbf{R})=\langle \mathbf{R}|n^X_i,n^Y_i,n^Z_i+1\rangle$), and therefore the overlap integrals are no longer equal. This allows us to have larger Ising couplings $\chi_{ij}$.

In addition to the interaction term, there are extra transverse and longitudinal fields generated by the interrogating laser. 
For blue sideband transitions, the single-particle Hamiltonian can be written as $H_{\mathrm{sp}}=\sum_i(\Omega_iS_i^x-(\delta-\hbar\omega) S_i^z)$, where $\Omega_i$ is the mode-dependent Rabi frequency, $\delta$ is the laser detuning from the carrier transition, and $\omega$ is the relevant trapping frequency.
Both $H_{\mathrm{sp}}$ and $H_{\mathrm{int}}$ (see Eq.(\ref{eq:xxz})) contribute to the dynamics in our XXZ simulator ($H_{\mathrm{XXZ}}=H_{\mathrm{sp}}+H_{\mathrm{int}}$), and the dynamics can be restricted to the fully symmetric Dicke manifold to the leading order. 
In this limit our model simplifies to the Lipkin-Meshkov-Glick (LMG) model \cite{lipkin1965}, 
\begin{equation}
H_{\mathrm{LMG}}=\chi S^zS^z+\Omega S^x-\tilde{\delta}S^z.
\label{eq:lmg}
\end{equation} 
Here, $\tilde{\delta}=\delta-\hbar\omega-B$ is the effective longitudinal field, $\chi, \Omega$ and $B$ are the thermal-averaged value of $\chi_{ij}, \Omega_{i}$ and $B_{i}$ respectively, and $S^{x,y,z}=\sum_iS_i^{x,y,z}$ are the collective spin operators. 

The LMG model features interesting spin dynamics, including a ferromagnetic to paramagnetic dynamical phase transition (DPT) \cite{muniz2020,zhang2017, Borish2020}. 
In general terms, a DPT is characterized by the existence of a critical point separating phases with distinct dynamical properties in many-body systems. 
The analog of thermodynamic order parameters is found in long-time average observables, which have a nonanalytic dependence on system parameters. 
To observe the DPT we initialize all the atoms in the $|\downarrow\rangle$ state, which is the ground state of LMG model when $\tilde{\delta}\rightarrow-\infty$, and then perform a sudden quench of the longitudinal field to its final value $\tilde{\delta}$. 

In this case, the DPT is signaled by a sharp change in behavior of the long-time average excitation fraction, $\overline{N_{\uparrow}}/N=\lim_{T\rightarrow\infty}\frac{1}{T}\int_0^TN_{\uparrow}(t)/N$, which serves as an order parameter and  distinguishes two dynamical phases (see Fig.~\ref{fig2}(a)): 
A dynamical ferromagnetic phase  characterized by $\overline{N_{\uparrow}}/N\approx 0$, where the vanishing excitation fraction persists even when the final longitudinal field $\tilde{\delta}$ is varied, 
and a dynamical paramagnetic phase, where $\overline{N_{\uparrow}}/N$ dynamically adjusts itself following the change of the final longitudinal field $\tilde{\delta}$. 

To analyze the DPT we derive mean-field equations of motion for the collective spin operators. They are given by
\begin{equation}
\frac{\mathrm{d}}{\mathrm{d}t}\mathbf{s}=\mathbf{b}\times\mathbf{s}, \quad \mathbf{b}=\Big(\Omega,0,N\chi s^z-\tilde{\delta}\Big),
\label{eq:mean}
\end{equation}
where $s^{x,y,z}=2\langle S^{x,y,z}\rangle/N$ are the normalized expectation values of collective spin operators. 
As shown in \cite{Note1}, Eq.(\ref{eq:mean}) can be further reduced to $(\dot{s}^z)^2/2+V(s^z)=0$ by eliminating $s^x$ and $s^y$, and we can relate the DPT with an abrupt change in the number of real roots of the effective potential $V(s^z)$ in this form. 
The abrupt change in $V(s^z)$ gives rise to distinct properties in spin dynamics shown in Fig.~\ref{fig2}(b): 
The ferromagnetic phase features small oscillations near south pole (trajectory C),
while the paramagnetic phase  exhibits large excursions that precess around the $x$ axis  (trajectory B). 
The DPT can also be tuned by varying the interaction strength as shown in Fig.~\ref{fig3}(d).
In the interaction dominant regime ($N\chi/\Omega>8\sqrt{3}/9$), the DPT generates a second order critical line (marked by the black solid line in Fig.~\ref{fig3}(d)) that distinguishes the two dynamical phases. On the other hand, the transition evolves into a smooth crossover region in the weakly interacting regime ($N\chi/\Omega<8\sqrt{3}/9$, below the black dashed line in Fig.~\ref{fig3}(d)), where instead the dynamics is   dominated by single particle Rabi flopping.

We experimentally realize the XXZ spin model in a cloud of $^{87}$Rb atoms, which is prepared at a temperature of $375(25)$~nK in a crossed-beam optical dipole trap with trapping frequencies of $143$~Hz, $21.5$~Hz and $171$~Hz.
This setting ensures the validity of the key approximations in our spin model, including the collisionless regime and a negligible number of doubly occupied modes (below $1.4\%$ for $10^5$ atoms) \cite{Note1}.
The atomic ensemble is initialized with a variable mean density $n$ from $0.46$ to $4.8\times10^{12}\mathrm{cm}^{-3}$ (atom number $N$ from $0.33$ to $3.4\times 10^5$), and the atom densities are calibrated by the collisional frequency shift of the carrier transition ($-0.48\,\mathrm{Hz}/10^{12}\mathrm{cm}^{-3}$ \cite{buning2011}).
To ensure an unperturbed cloud temperature for different atom densities, an adjustable spin rotation is performed, which partially transfers atoms from the $|\uparrow\rangle$ to the $|\downarrow\rangle$ state and a subsequent removal of the $|\uparrow\rangle$ atoms.
The coherent drive between two clock states with resolved motional levels is realized by two copropagating Raman beams focused into the atomic cloud with a $39\,\mu$m beam waist. The beams are offset from the trap center in order to drive the first-order motional sidebands \cite{Note1}. 
The typical Rabi spectrum of our system is depicted in Fig.~\ref{fig1}(b). 
Here we focus on the strongest blue sideband at $\omega/2\pi=171$~Hz.
Considering the mean Ising couplings ($N\chi/h\approx 4.63\,\mathrm{Hz}/10^{12}\mathrm{cm}^{-3}$) and the mean Rabi frequency ($\Omega/h\approx 0.56$~Hz) for this sideband, our XXZ simulator lies in the interaction dominant regime, where the mentioned DPT is predicted to occur. 
Instead of direct measurements of the long-time-averaged excitation fraction, which is inevitably limited by technical challenges (e.g. collisional dephasing and atom loss), the order parameter $\overline{N_{\uparrow}}/N$ is estimated by measuring the excitation fraction at a probe time $t_f=0.5$s for fixed Rabi frequency.
The entire phase diagram is then obtained by scanning the two-photon detuning $\delta$ in $0.8$~Hz steps and by varying interactions using different atom densities.

The recorded asymmetric lineshapes for different atom densities are shown in Fig.~\ref{fig3}(a-c, e-f), which is in good agreement with the mean-field theoretical predictions by $H_{\mathrm{XXZ}}$ \cite{Note1}.
In Fig.~\ref{fig3}(e), we also compare the experimental observation with the order parameter $\overline{N_{\uparrow}}/N$ (green dot-dashed line) predicted by $H_{\mathrm{LMG}}$ (see Eq.(\ref{eq:lmg})). 
We find that the recorded lineshape captures the two dynamical phases in the LMG model: if we increase the two-photon detuning $\delta$, the slow increase of $N_{\uparrow}/N$ below resonance indicates the paramagnetic phase, while the sharp change back to $N_{\uparrow}/N\approx 0$ above resonance indicates the ferromagnetic phase. 
Compared to the critical behavior of $\overline{N_{\uparrow}}/N$ in the LMG model, the recorded lineshapes are broadened by the inhomogeneous couplings but retain the sharp features associated with the DPT.
The inhomogeneities also lead to modifications of effective interaction strength in experiment compared to the LMG model, which can be accounted for by scaling $\chi$ by a factor of $0.56$.
By interpreting the experimentally observed resonant frequencies (obtained from maximal population transfer) as a signature of the critical point of the DPT, we reconstruct the phase boundary between these two dynamical phases (see Fig.~\ref{fig3}(d)), which agrees with the theoretical prediction.

\begin{figure}[t]
    \includegraphics{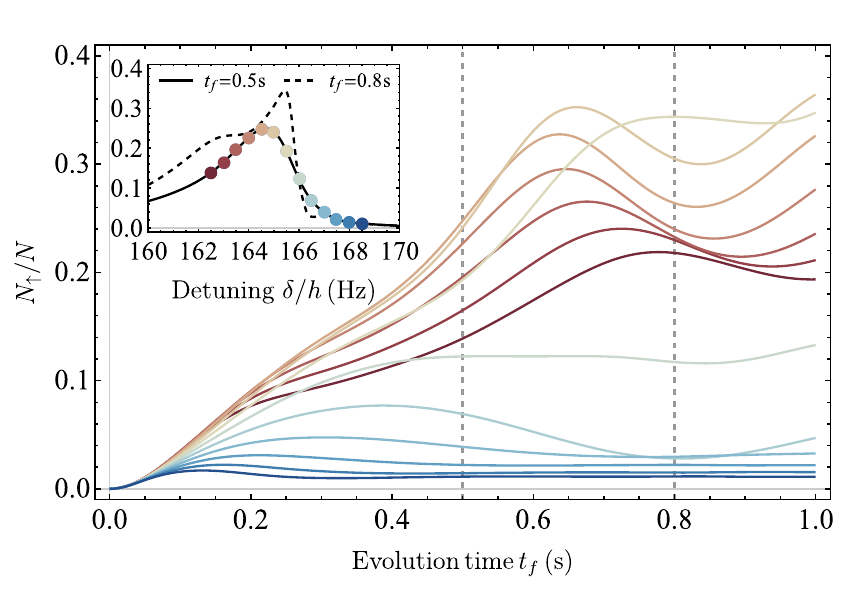}
    \caption{\label{fig4}Numerical simulation of mean-field evolution under $H_{\mathrm{XXZ}}$ with mean atom density $n=3.2\times 10^{12}\mathrm{cm}^{-3}$. The lines with a color gradient from red to blue show the dynamical behavior from red to blue detuning with $0.5$~Hz frequency steps. A sharp change in the excited atom fraction can be observed as the system approaches the critical point. The inset compares two lineshapes taken at different evolution times.}
\end{figure}

To further verify the existence of a DPT with inhomogeneous couplings, we present numerical simulations of mean-field evolution under $H_{\mathrm{XXZ}}$ with mean atom density $n=3.2\times10^{12}\mathrm{cm}^{-3}$ in Fig.~\ref{fig4}. 
As the detuning is scanned from below to above the critical point (marked by a color gradient from red to blue), the excitation fraction $N_{\uparrow}(t)/N$ features a sharp change in dynamical behavior at the critical detuning, validating the existence of a DPT under our experimental conditions. 
Compared to the LMG model, we observe damping in the oscillation amplitude of excitation fraction for the inhomogeneous case. 
To understand the role of the damping, in the inset of Fig.~\ref{fig4} we compare the lineshapes at evolution times of $t_f=0.5$s (see also Fig.~\ref{fig3}) and $t_f=0.8$s. 
Although we see variations in the lineshapes computed at these two evolution times (the latter is sharper than the former), both of them consistently display clear signatures of the DPT up to a $1$~Hz shift in resonant frequency, which nevertheless lies within the experimental error bars. This analysis justifies the use of  $N_{\uparrow}/N$ evaluated at $t_f=0.5$s as a good proxy for the long-time-averaged order parameter.

In summary, we have demonstrated the use of motional sidebands in trapped bosonic gases as a tool to simulate long-range XXZ spin models.
A practical application of the demonstrated sideband protocol is the dynamical generation of spin squeezing, a well known feature of the LMG model \cite{ma2011} which makes it useful for enhanced sensing.
Although further control of inhomogeneties will be required to observe squeezing in the current setup, we expect spin squeezing can be in reach in the next generation of experiments \cite{Note1}.
Moreover, we expect our protocol can be feasibly implemented in a wide range of experiments, including atomic systems in optical lattices. 
In these systems the SU(2) symmetry of superexchange interactions could be broken into a XXZ spin model via  motional sideband spectroscopy, thanks to the larger tunneling rates of excited bands.

\begin{acknowledgments}
We thank Itamar Kimchi and Diego Barberena for useful discussions. 
The theoretical work is supported by the AFOSR grant FA9550-18-1-0319, by the DARPA (funded via ARO) grant W911NF-16-1-0576, the ARO single investigator award W911NF-19-1-0210, the NSF PHY1820885, NSF JILA-PFC PHY-1734006 grants, and by NIST. 
The experimental work is funded by the Deutsche Forschungsgemeinschaft (DFG, German Research Foundation) under Germany's Excellence Strategy -- EXC-2123 QuantumFrontiers -- 390837967, and through CRC 1227 (DQ-mat), project A02.
\end{acknowledgments}

\end{document}


\title{Simulation of XXZ Spin Models using Sideband Transitions in Trapped Bosonic Gases: Supplemental Materials}
\author{Anjun Chu$^{1,2}$, Johannes Will$^{3}$, Jan Arlt$^{4}$, Carsten Klempt$^{3}$, and Ana Maria Rey$^{1,2}$}
\affiliation{$^{1}$JILA, NIST and Department of Physics, University of Colorado, Boulder, CO 80309, USA\\
$^{2}$Center for Theory of Quantum Matter, University of Colorado, Boulder, CO 80309, USA\\
$^{3}$Institut f\"{u}r Quantenoptik, Leibniz Universit\"{a}t Hannover, Welfengarten 1, D-30167 Hannover, Germany\\
$^{4}$Institut for Fysik og Astronomi, Aarhus Universitet, Ny Munkegade 120, DK-8000 Aarhus C, Denmark}

\maketitle

\section{Experimental realization}
\begin{figure*}[ht!]
    \includegraphics{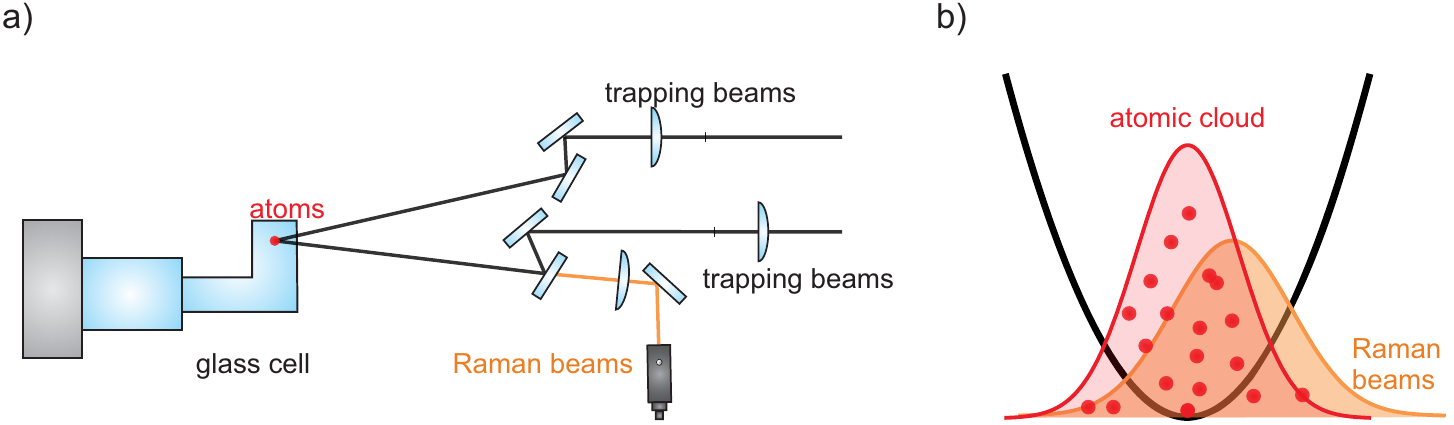}
    \caption{\label{sfig00}(a) Experimental implementation. The optical dipole potential is formed by two laser beams (black) which intersect with an angle of $18^\circ$ in the experimental chamber. The Raman laser beams (orange) are directed onto the atomic cloud along one of these beams. (b) To address the motional sidebands an asymmetric  coupling field is required, which is realized by introducing a small offset between the Raman laser focus and the atomic cloud.}
\end{figure*}

To initiate our experiments, ultracold $^{87}$Rb atoms are captured in a magneto-optical trap loaded from a background gas. The captured atoms are subsequently transported to a vacuum region with reduced pressure, where they are cooled by forced evaporative cooling. Initial cooling is performed in a hybrid trap that combines a magnetic quadrupole trap with an optical dipole potential~\cite{Lin2009}. The final temperatures are achieved by evaporation in a pure optical dipole trap. This trap is formed by two laser beams at a wavelength of $1064$~nm,  which intersect at an angle of $18^\circ$ as shown in Fig.~\ref{sfig00}(a). The two beams have waists of $60$ and $75\,\mu$m, which provide a nearly harmonic trapping potential with trapping frequencies of $143$~Hz, $21.5$~Hz and $171$~Hz at the chosen laser power.

\begin{figure*}[ht!]
    \includegraphics[width=0.8\textwidth]{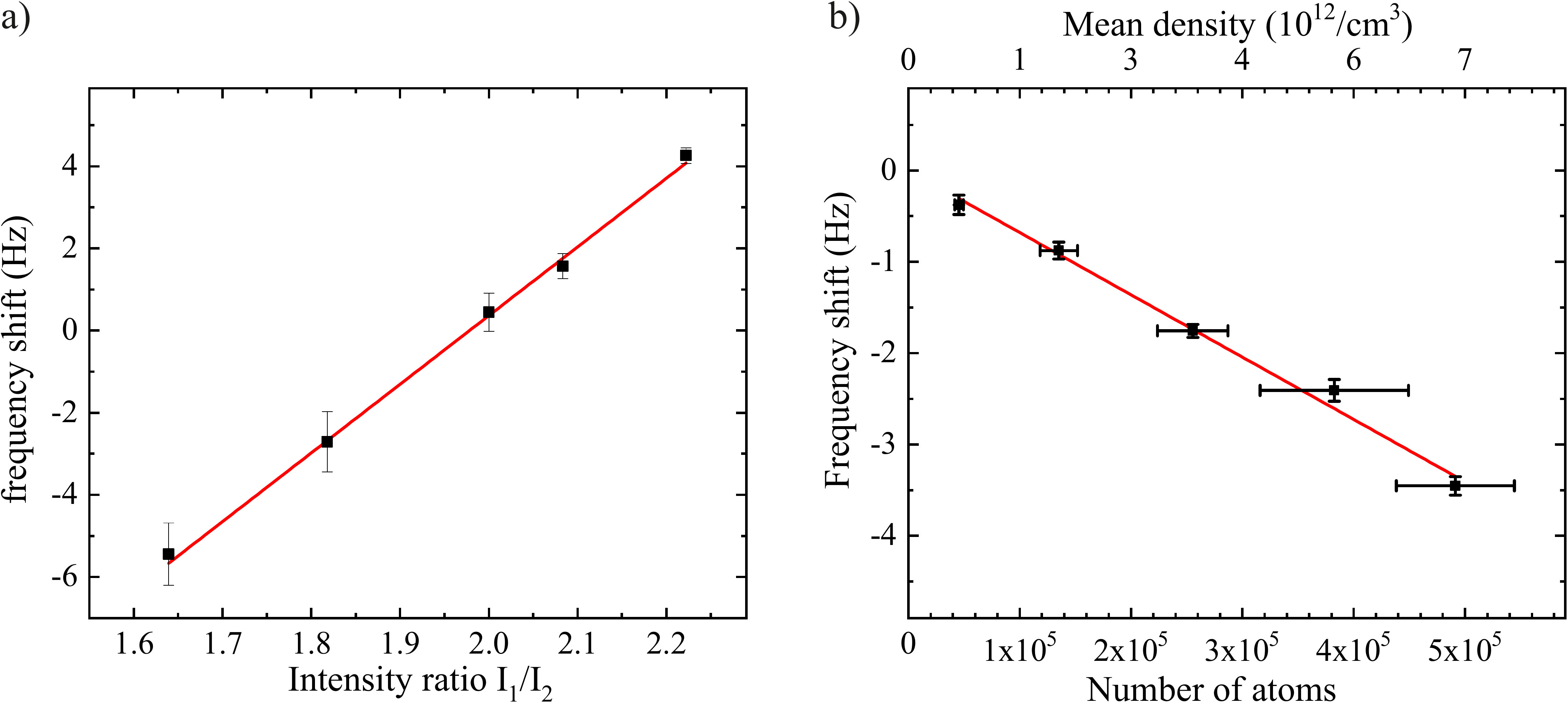}
    \caption{\label{sfig01} (a) Frequency shift observed using the Raman transition relative to the frequency obtained by using the corresponding microwave transition. The shift vanishes for an intensity ratio of $I_1/I_2=1.98$. (b) Calibration of the mean density. The measured frequency of the carrier transition is shown as a function of the recorded number of atoms. Based on the known density-dependent frequency shift for Rb atoms \cite{buning2011}, the mean density of the ensemble is obtained.}
\end{figure*}

The procedure outlined above results in an ensemble of $4 \times 10^5$ atoms in the  $\ket{F\!=\!2, m_F\!=\!2}$ state at a temperature of $375(25)$~nK. Subsequently, the atoms are transferred to the $\ket{\uparrow}=\ket{F\!=\!2, m_F\!=\!0}$ state by a radio-frequency rapid adiabatic passage. In a final step, the atomic ensemble is initialized with a variable mean density $n$ between $0.46$ and $4.8\times10^{12}\mathrm{cm}^{-3}$ in the $\ket{\downarrow}=\ket{F\!=\!1, m_F\!=\!0}$ state by a microwave Rabi pulse of variable duration. The remaining atoms in the $\ket{\uparrow}$ state are removed with a resonant light pulse on an optical transition.
Importantly, this method allows for the preparation of a variable density at constant temperature. The density of these ensembles is calibrated by performing microwave-based Ramsey interferometry on the clock transition (Fig.~\ref{sfig01}(b)) and by recording the density-dependent frequency shift of the carrier transition ($-0.48\,\mathrm{Hz}/10^{12}\mathrm{cm}^{-3}$ \cite{buning2011}). 
Here we provide a list that connects the mean atom density $n$ and the corresponding atom number $N$ we used in the DPT experiment:
\begin{center}
\renewcommand\arraystretch{1.5}
\begin{tabular}{|c||c|c|c|c|c|}
\hline
Mean atom density $n\;(\mathrm{cm}^{-3})$ & $0.46\times 10^{12}$ & $0.98\times 10^{12}$ & $2.0\times 10^{12}$ & $3.2\times 10^{12}$ & $4.8\times 10^{12}$\\
\hline
Atom number $N$ & $0.33\times 10^5$ & $0.69\times 10^5$ & $1.4\times 10^5$ & $2.2\times 10^5$ & $3.4\times 10^5$\\
\hline
\end{tabular}
\end{center}

The inhomogeneous coupling field between the two clock states $\ket{\downarrow}$ and $\ket{\uparrow}$ is realized by using two copropagating Raman beams which are derived from two phase-locked diode lasers. The Raman beams are focused onto the atomic cloud with a beam waist of $39\,\mu$m. In principle, these Raman beams can lead to a differential shift of the clock states and thus fluctuations in the Raman beam power may lead to a significant broadening of the spectroscopic signal. To avoid this effect, a specific relative intensity of the Raman beams can be chosen, which reduces the light shift~\cite{Kaplan2002}. Figure~\ref{sfig01}(a)) shows the experimental determination of the optimal relative intensity based on the comparison of spectroscopy on the clock transition with the Raman system and with microwave radiation. In the experiments a relative intensity $I_1/I_2=1.98$ was chosen to avoid differential shifts. 

The first-order motional sidebands can only be addressed with a coupling field that is asymmetric with respect to the trapping potential. This asymmetry is realized by shifting the Raman beams compared to the center position of the atomic cloud as shown in Fig.~\ref{sfig00}(b) and leads to a spatial inhomogeneity of the coupling. 

The spectroscopic signals shown in of Fig.~1 and Fig.~2 (main text) are obtained by applying Raman pulses with a duration of $500$~ms for detunings between $\approx-200$~Hz and $\approx+200$~Hz of the two Raman laser beams. In Fig.~1 (main text), the carrier and all six sidebands are well resolved and, in addition, a higher order sideband is visible at $\approx192$~Hz. Compared to previous work~\cite{Maineult2012}, the small Fourier width of these pulses allow for a full resolution of the sideband transitions.

At the end of each experimental sequence, the trap is switched off to allow for ballistic expansion and Stern-Gerlach separation of the atoms in the two clock states. The number of atoms in both states, $N_\uparrow$ and $N_\downarrow$,
and their temperature are detected by simultaneous absorption imaging, calibrated according to Ref.~\cite{Reinaudi2007}.

\section{Spin Model and Mean-field Dynamics }

In the main text we show that long-range XXZ spin models describe trapped bosonic gases interacting via purely contact interactions. Here we discuss the various parameters of the XXZ spin models and derive the corresponding mean-field equations of motion. We use them to calculate associated Rabi lineshapes.

Recall the Hamiltonian $H_{\mathrm{XXZ}}$ defined in main text,
\begin{equation}
    H_{\mathrm{XXZ}}=\sum_{ij}J_{ij}\mathbf{S}_i\cdot\mathbf{S}_j+\sum_{ij}\chi_{ij}S_i^zS_j^z+\sum_i\Omega_iS_i^x-\sum_i(\delta-B_i)S_i^z.
    \label{eq:spin}
\end{equation} 
This Hamiltonian describes the  spin dynamics of thermal bosonic gases in the collisionless regime. By adequately mapping  the harmonic trap eigenmodes to lattice sites in mode space, we can understand the spin dynamics for the carrier transition, the  blue sideband transition, as well as the red sideband transition. The definitions of the two spin states in lattice site $i$ for all these cases are as follows:
\begin{itemize}
    \item Carrier transition: $|\Uparrow_i\rangle=|\uparrow;n_i^X,n_i^Y,n_i^Z\rangle$, $|\Downarrow_i\rangle=|\downarrow;n_i^X,n_i^Y,n_i^Z\rangle$
    \item Blue sideband transition ($\hat{Z}$ direction): $|\Uparrow_i\rangle=|\uparrow;n_i^X,n_i^Y,n_i^Z+1\rangle$, $|\Downarrow_i\rangle=|\downarrow;n_i^X,n_i^Y,n_i^Z\rangle$
    \item Red sideband transition ($\hat{Z}$ direction): $|\Uparrow_i\rangle=|\uparrow;n_i^X,n_i^Y,n_i^Z-1\rangle$, $|\Downarrow_i\rangle=|\downarrow;n_i^X,n_i^Y,n_i^Z\rangle$
\end{itemize}
Here we will use the convention of capital letters to denote spatial coordinates to distinguish them from coordinates in spin space denoted by lowercase letters. We denote the wave functions associated with $|\Uparrow_i\rangle$ and $|\Downarrow_i\rangle$ states respectively as $\phi_i^{\Uparrow}(\mathbf{R})$ and $\phi_i^{\Downarrow}(\mathbf{R})$. To avoid confusion, we define $\delta$ as the laser detuning from the carrier transition, and this convention is also used in the main text. For the blue sideband transition, we replace $\delta$ by $\delta-\hbar\omega$, where $\omega$ is the relevant trapping frequency; while for red sideband transition, we replace $\delta$ by $\delta+\hbar\omega$.

\begin{figure*}[t]
    \includegraphics{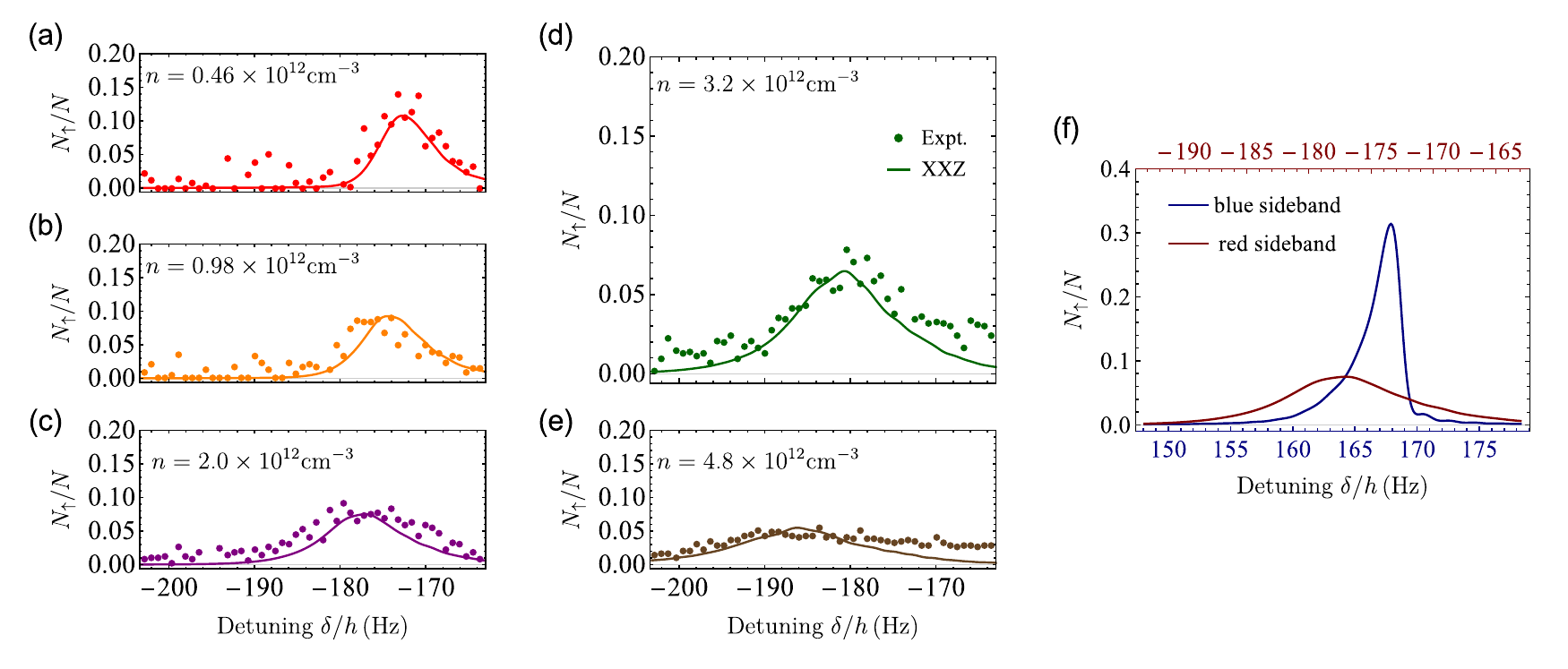}
    \caption{\label{sfig1}(a-e) The Rabi lineshapes for $171$~Hz red sideband with mean atom density $n=\{0.46,0.98,2.0,3.2,4.8\}\times 10^{12}\mathrm{cm}^{-3}$. The solid points denote experimental data, the solid lines denote mean-field theoretical predictions by $H_{\mathrm{XXZ}}$ (see text). (f) The comparison of theoretical Rabi lineshapes between blue sideband and red sideband with mean density $n=2.0\times 10^{12}\mathrm{cm}^{-3}$, and the difference can be understood as the effect of the anharmonicity of the optical dipole trap (see text).}
\end{figure*}

As we discuss in the main text, the key approximations in our spin model are the collisionless regime (trapping frequency is much larger than the interaction strength) as well as a negligible number of doubly occupied modes. Our experimental system is a thermal cloud of $^{87}$Rb atoms prepared at a temperature $T=375(25)$~nK in a 3D harmonic trap with trapping frequencies $\omega_X/2\pi=143$~Hz, $\omega_Y/2\pi=21.5$~Hz, $\omega_Z/2\pi=171$~Hz. Based on the numerical calculation described below, we find $N\chi/h\approx 4.63\mathrm{Hz/10^{12}cm^{-3}}$ for blue sideband transition in the $\hat{Z}$ direction, which demonstrates the validity of collisionless regime. For the number of doubly occupied modes, we evaluate the quantity $\sum_i\langle \hat{n}_i(\hat{n}_i-1)\rangle/2N$, where $\hat{n}_i=\hat{a}^{\dag}_i\hat{a}_i$, and we are summing over all possible eigenmodes. This quantity is only non-zero when there is more than one atom per mode. Using Wick's theorem, we have
\begin{equation}
\frac{1}{2N}\sum_i\langle\hat{n}_i(\hat{n}_i-1)\rangle=\frac{1}{2N}\sum_i\langle \hat{a}^{\dag}_i\hat{a}^{\dag}_i\hat{a}_i\hat{a}_i\rangle=\frac{1}{N}\sum_i\langle \hat{a}^{\dag}_i\hat{a}_i\rangle\langle \hat{a}^{\dag}_i\hat{a}_i\rangle=\frac{1}{N}\sum_i\langle \hat{n}_i\rangle^2
\end{equation}
Then we can estimate the average occupation number of each eigenmode by a Boltzmann distribution, $\langle \hat{n}_i\rangle/N=\exp[-(n^X\omega_X+n^Y\omega_Y+n^Z\omega_Z)\hbar/k_BT]/Z$, in which the partition function $Z\approx (k_BT/\hbar\bar{\omega})^3$. Then we get
\begin{equation}
\frac{1}{2N}\sum_i\langle \hat{n}_i(\hat{n}_i-1)\rangle\approx N\bigg(\frac{\hbar\bar{\omega}}{2k_BT}\bigg)^3
\end{equation}
For $N=1\times 10^5$ atoms, we obtain that the number of occupied eigenmodes with two atoms or more is $N(\hbar\bar{\omega}/2k_BT)^3\sim 1.4\%$. This result leads to our conclusion that the number of lattice sites occupied by more than one atoms is very small in our experiment. 

For the bosonic gas of interest, the Heisenberg couplings $J_{ij}$ and the Ising couplings $\chi_{ij}$ are purely generated by contact interactions, and can be written in terms of the various scattering lengths and overlap integrals of harmonic trap eigenmodes,
\begin{equation}
    J_{ij}=\frac{4\pi\hbar^2 a_{\uparrow\downarrow}V_{ij}^{\mathrm{ex}}}{m}, \quad\chi_{ij}=\frac{4\pi\hbar^2(V_{ij}^{\Uparrow\Uparrow}a_{\uparrow\uparrow}+V_{ij}^{\Downarrow\Downarrow}a_{\downarrow\downarrow}-V_{ij}^{\Uparrow\Downarrow}a_{\uparrow\downarrow}-V_{ij}^{\mathrm{ex}}a_{\uparrow\downarrow})}{m},
\end{equation}
where
\begin{equation}
    V_{ij}^{\alpha\beta}=\int\mathrm{d}^3\mathbf{R}[\phi_i^{\alpha}(\mathbf{R})]^2[\phi_j^{\beta}(\mathbf{R})]^2, \quad V_{ij}^{\mathrm{ex}}=\int\mathrm{d}^3\mathbf{R}\phi_i^{\Uparrow}(\mathbf{R})\phi_i^{\Downarrow}(\mathbf{R})\phi_j^{\Uparrow}(\mathbf{R})\phi_j^{\Downarrow}(\mathbf{R}).
\end{equation}.

Besides spin-spin interaction, the Raman laser couples $|\Uparrow_i\rangle$ and $|\Downarrow_i\rangle$ states via a intermediate state, and the effective Rabi frequencies are given by 
\begin{equation}
    \Omega_i=\Omega_0\int\mathrm{d}^3\mathbf{R}\exp\bigg[-\frac{2(\mathbf{R}_{\perp}-\mathbf{R}_{\perp,0})^2}{w^2}\bigg]\phi_i^{\Uparrow}(\mathbf{R})\phi_i^{\Downarrow}(\mathbf{R}).
\end{equation}
Here, $\Omega_0$ is the bare Rabi frequency determined by the Rabi couplings and detunings to the intermediate state of the two Raman beams, $\mathbf{R}_{\perp}$ is perpendicular to the propagating direction of Raman beams, $\mathbf{R}_{\perp,0}$ is the offset from trap center, and $w$ is the Gaussian beam radius at the position of atomic cloud. Due to the copropagating geometry of Raman beams used in the experiment, the momentum kicks of Raman beams can be ignored.

The different inhomogeneous longitudinal fields $B_i=B_i^{(1)}+B_i^{(2)}+B_i^{(3)}$ come from the interplay between contact interaction ($B_i^{(1)}$), the differential frequency shift generated by the optical trap and the magnetic curvature ($B_i^{(2)}$), and the anharmonicity in optical trap ($B_i^{(3)}$). We already discuss $B_i^{(1)}$ in the main text. It is given by
\begin{equation}
    B_i^{(1)}=\frac{4\pi\hbar^2}{m}\sum_{j\neq i}(V_{ij}^{\Uparrow\Uparrow}a_{\uparrow\uparrow}-V_{ij}^{\Downarrow\Downarrow}a_{\downarrow\downarrow}).
\end{equation}

The different trapping frequency experienced by the $|\uparrow\rangle$ and $|\downarrow\rangle$ atoms give an additional differential frequency shift in mode space. If we defined $\Delta \omega_{X,Y,Z}=\omega_{X,Y,Z}^{\uparrow}-\omega_{X,Y,Z}^{\downarrow}$, we can express $B_i^{(2)}$ as
\begin{equation}
    B_i^{(2)}=\bigg(n_i^X+\frac{1}{2}\bigg)\hbar\Delta\omega_X+\bigg(n_i^Y+\frac{1}{2}\bigg)\hbar\Delta\omega_Y+\bigg(n_i^Z+\frac{1}{2}\bigg)\hbar\Delta\omega_Z.
\end{equation}

Moreover, the actual Gaussian shape of the laser beams that make the dipole trap introduces corrections beyond the leading order harmonic trapping potential $U(\mathbf{R})=\frac{1}{2}m(\omega_X^2X^2+\omega_Y^2Y^2+\omega_Z^2Z^2)$. These corrections generate an additional anharmonic potential $\Delta U(\mathbf{R})=-\frac{1}{2}m(\gamma_{XX}^2X^4+\gamma_{YY}^2Y^4+\gamma_{ZZ}^2Z^4+\gamma_{XY}^2X^2Y^2+\gamma_{XZ}^2X^2Z^2+\gamma_{YZ}^2Y^2Z^2)$. In first-order perturbation, this term gives rise to a shift of harmonic oscillator levels, which leads to a small change of atom density. As the shift generated by the anharmonicity is mode-dependent, it generates an extra longitudinal field that should be taken in consideration for sideband transitions. This field has a sign difference for the  blue sideband ($B_i^{(3)b}$) and red sideband ($B_i^{(3)r}$). Here we use the sideband transitions in $\hat{Z}$ direction as an example,
\begin{equation}
    B_i^{(3)b}=-\frac{\hbar}{2\omega_Z}\bigg[3\gamma_{ZZ}^2(a_{\mathrm{ho}}^Z)^2(n_i^Z+1)+\gamma_{YZ}^2(a_{\mathrm{ho}}^Y)^2(n_i^Y+1/2)+\gamma_{XZ}^2(a_{\mathrm{ho}}^X)^2(n_i^X+1/2)\bigg],
\end{equation}
\begin{equation}
    B_i^{(3)r}=\frac{\hbar}{2\omega_Z}\bigg[3\gamma_{ZZ}^2(a_{\mathrm{ho}}^Z)^2n_i^Z+\gamma_{YZ}^2(a_{\mathrm{ho}}^Y)^2(n_i^Y+1/2)+\gamma_{XZ}^2(a_{\mathrm{ho}}^X)^2(n_i^X+1/2)\bigg],
\end{equation}
where $a_{\mathrm{ho}}^{X,Y,Z}=\sqrt{\hbar/m\omega_{X,Y,Z}}$ is the harmonic oscillator length. Sideband transitions in other directions can be treated in a similar way.

For carrier transition, due to the negligible Ising couplings $\chi_{ij}$, our XXZ spin model can be simplified to the Heisenberg model. Given that the transverse field $\Omega_i$ and longitudinal field $B_i$ are small compared to the Heisenberg couplings $J_{ij}$, we can restrict the spin model in the Dicke manifold, which gives $H_{\mathrm{carrier}}\approx \Omega S^x-(\delta-B)S^z$, where $\Omega$ is the mean Rabi frequency for carrier transition, and $B$ is the thermal-averaged value of $B_i$. We understand $B$ as the frequency shift of carrier transition, which can be evaluated analytically in the large-$N$ limit,
\begin{equation}
    B=\frac{4\pi\hbar^2(a_{\uparrow\uparrow}-a_{\downarrow\downarrow})n}{m}+k_BT\bigg(\frac{\Delta\omega_X}{\omega_X}+\frac{\Delta\omega_Y}{\omega_Y}+\frac{\Delta\omega_Z}{\omega_Z}\bigg),
\end{equation}
where $n=N(m\bar{\omega}^2/4\pi k_BT)^{3/2}$ is the mean atom density in harmonic trap with atom number $N$, and $\bar{\omega}=(\omega_X\omega_Y\omega_Z)^{1/3}$. The density-dependent part in $B$ agrees with the density-dependent clock shift $-0.48~\mathrm{Hz/10^{12}cm^{-3}}$ observed in previous experiment \cite{buning2011}. In our experiment, we use this known value of density-dependent shift of the carrier transition to calibrate the atom density.

For sideband transitions, the Ising couplings $\chi_{ij}$ become larger. We use a  mean-field approximation, which neglects the quantum correlation between different spins, $\langle S_i^{\mu}S_j^{\mu'}\rangle\approx\langle S_i^{\mu}\rangle\langle S^{\mu'}_j\rangle$ ($\mu,\mu'=x,y,z$), to derive Heisenberg equations of $S_i^{x,y,z}$ in our XXZ spin model (see Eq.(\ref{eq:spin})). The mean-field equations we get are the following ones:
\begin{equation}
    \begin{gathered}
    \frac{\mathrm{d}}{\mathrm{d}t}\langle S_j^{x}\rangle=2\sum_i\bigg[J_{ij}\langle S_i^y\rangle\langle S_j^z\rangle-(J_{ij}+\chi_{ij})\langle S_i^z\rangle \langle S_j^y\rangle\bigg]+(\delta-B_j)\langle S_j^y\rangle,\\
    \frac{\mathrm{d}}{\mathrm{d}t}S_j^{y}=2\sum_{i}\bigg[(J_{ij}+\chi_{ij})\langle S_i^z\rangle \langle S_j^x\rangle-J_{ij}\langle S_i^x\rangle \langle S_j^z\rangle\bigg]-(\delta-B_j)\langle S_j^x\rangle-\Omega_j\langle S_j^z\rangle,\\
    \frac{\mathrm{d}}{\mathrm{d}t}\langle S_j^{z}\rangle=2\sum_{i}J_{ij}\bigg[\langle S_i^x\rangle\langle S_j^y\rangle-\langle S_i^y\rangle \langle S_j^x\rangle\bigg]+\Omega_j\langle S_j^y\rangle.\\
    \end{gathered}
    \label{eq:mean}
\end{equation}
We solve Eq.(\ref{eq:mean}) numerically, with random sampling of motional states drawn from a Boltzmann distribution. As it is computationally difficult  to solve the equations above for $\sim 10^5$ atoms, instead we use $N_{\mathrm{th}}=1000$ and scale the transverse and longitudinal field from the one in the experiment by a factor $N_{\mathrm{th}}/N_{\mathrm{exp}}$. We also allow an overall scaling factor $\eta$ of the atomic density to take both  finite-size effects and the anharmonicities into account. The thermal-averaged sideband spectrum agrees well with our experimental measurements, when  the overall scaling factor is set to $\eta=0.72$ for both blue sideband and red sideband. The Rabi spectrum of blue sideband transition is discussed in the main text (see Fig.~3(a-c,e-f)), and the Rabi spectrum of red sideband transition is depicted in Fig.~\ref{sfig1}(a-e). We also compare the theoretical Rabi lineshapes for blue and red sideband for mean atom density $n=2.0\times 10^{12}\mathrm{cm}^{-3}$ in Fig.~\ref{sfig1}(f), which shows a significant suppression of red sideband. As the temperature of our system is above quantum degeneracy, the ground state concentration is not a reasonable explanation. Instead, the difference between the blue and red sidebands comes from a sign difference between $B_i^{(3)b}$ and $B_i^{(3)r}$, generated by anharmonicity. Because of this sign difference, $B_i^{(3)b}$ partially cancels the inhomogeneity in the longitudinal fields, while $B_i^{(3)r}$ increases the inhomogeneity. Similar phenomenon was also observed in Ref.~\cite{allard2016}.

\section{Dynamical phase diagram and critical behavior}
In the main text we discuss about the ferromagnetic to paramagentic dynamical phase transition (DPT) in the Lipkin-Meshkov-Glick (LMG) model (see Eq.(3) main text), a collective XXZ model plus additional transverse and longitudinal field. Here we elaborate on the calculation of the dynamical phase diagram and the associated critical points, following the procedure discussed in Ref.~\cite{muniz2020}. The mean-field equations of the LMG model can be written in terms of normalized expectation value of total spin operators $s^{x,y,z}=2\langle S^{x,y,z}\rangle/N$ as follows,
\begin{equation}
    \begin{gathered}
    \frac{\mathrm{d}}{\mathrm{d}t}s^x=-N\chi s^zs^y+\tilde{\delta}s^y,\\
    \frac{\mathrm{d}}{\mathrm{d}t}s^y=N\chi s^zs^x-\tilde{\delta}s^x-\Omega s^z,\\
    \frac{\mathrm{d}}{\mathrm{d}t}s^z=\Omega s^y.\\
    \end{gathered}
    \label{eq:dpt1}
\end{equation}

Using both   energy conservation in $H_{\mathrm{LMG}}$, for an initial state with $s^z=-1, \,s^x=s^y=0$, as well as the identity $(S^x)^2+(S^y)^2+(S^z)^2=\big(\frac{N}{2}+1\big)\frac{N}{2}$ in large-$N$ limit, the mean-field variables satisfy the following two conservation relations:
\begin{equation}
    \frac{N\chi}{2}s^zs^z-\tilde{\delta}s^z+\Omega s_x=\frac{N\chi}{2}+\tilde{\delta},
    \label{eq:dpt2}
\end{equation}
\begin{equation}
    (s^x)^2+(s^y)^2+(s^z)^2=1.
    \label{eq:dpt3}
\end{equation}

Combining these three equations (Eq.(\ref{eq:dpt1})-(\ref{eq:dpt3})), we can eliminate $s^x$ and $s^y$, and obtain the following differential equation for $s^z$,
\begin{equation}
    \frac{1}{2}\bigg(\frac{\mathrm{d}}{\mathrm{d}t}s^z\bigg)^2+V(s^z)=0,
    \label{eq:dpt4}
\end{equation}
where
\begin{equation}
    V(s^z)=(s^z+1)\bigg\{\frac{(N\chi)^2}{8}(s^z)^3-\bigg[\frac{(N\chi)^2}{8}+\frac{N\chi\tilde{\delta}}{2}\bigg](s^z)^2+\bigg[\frac{\tilde{\delta}^2+\Omega^2}{2}-\frac{(N\chi)^2}{8}\bigg]s^z+\bigg[\frac{\tilde{\delta}^2-\Omega^2}{2}+\frac{N\chi\tilde{\delta}}{2}+\frac{(N\chi)^2}{8}\bigg]\bigg\}.
\end{equation}

\begin{figure*}[t]
    \includegraphics{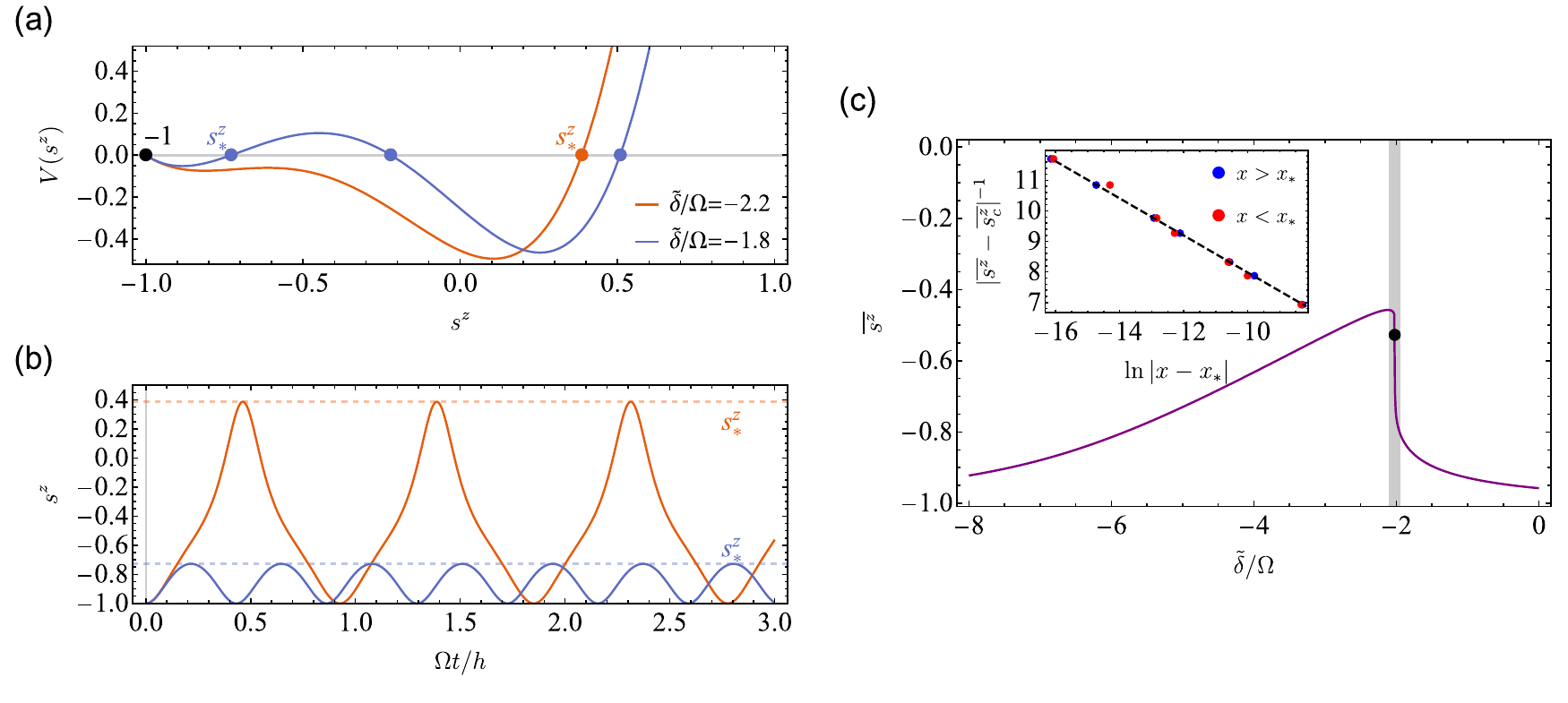}
    \caption{\label{sfig2}(a) The effective potential $V(s^z)$ with $N\chi/\Omega=5$. In the case of $\tilde{\delta}/\Omega=-2.2$, $V(s^z)$ has two real roots; In the case of $\tilde{\delta}/\Omega=-1.8$, $V(s^z)$ has four real roots. The nearest turnover point is labelled by $s^z_{*}$, and the jump of $s^z_{*}$ indicates dynamical phase transition (see text). (b) The mean-field dynamics of LMG model with $N\chi/\Omega=5$ and $\tilde{\delta}/\Omega=-2.2,-1.8$, which shows a sharp change of mean-field dynamical behavior. The choice of color for the lines is the same as (a), and the dashed lines mark the nearest turnover point $s^z_{*}$. (c) The long-time average value $\overline{s^z}$ with $N\chi/\Omega=5$, and the critical point is labelled by black circle. The shaded area is the region close to the critical point, which is shown in details in the inset with logarithmic non-analyticity at the critical point (see text).}
\end{figure*}

We interpret Eq.(\ref{eq:dpt4}) as the Hamiltonian of a classical particle with position $s^z$ moving in the effective potential $V(s^z)$. The condition  $V(s^z)=0$ determines  the turning points of $s^z$. Since $V(-1)=0$, $V'(-1)=-1$, $V(1)=\tilde{\delta}^2$, this effective potential should have at least two real roots in $[-1,1]$, and we consider these roots as physical turning points. So the dynamics of $s^z$ can be understood as the oscillations between $-1$ and the nearest turnover point $s^z_{*}$. Imagine that we start from a $V(s^z)$ with two real roots, and continuously tune the parameters of $V(s^z)$ so that two new real roots appear in between, a jump of the nearest turning point $s^z_{*}$ should occur in this process (see Fig.~\ref{sfig2}(a-b)). This abrupt change in behavior is what sets the dynamical phase transition.

To count the number of roots in $V(s^z)$, we can factor out the known root $s^z=-1$, and then consider the discriminant $\Delta=18abcd-4b^3d+b^2c^2-4ac^3-27a^2d^2$ of cubic equation $ax^3+bx^2+cx+d=0$. If $\Delta>0$, the cubic has three distinct real roots; if $\Delta<0$, the cubic has one real root. So $\Delta=0$ captures the critical point of the DPT. We focus on the parameter regime where $N\chi>0$ with a fixed positive $\Omega$, and define $y=N\chi/\Omega$, $x=\tilde{\delta}/\Omega$. In terms of these variables  the phase boundary plotted in the main text (see Fig.~2(d)) is given by
\begin{equation}
    y_{*}=\frac{1}{12x_{*}}\bigg[1-12x_{*}^2-\bigg(5832x_{*}^4+540x_{*}^2-1+24x_{*}\sqrt{3(27x_{*}^2-1)^3}\bigg)^{1/3}-\bigg(5832x_{*}^4+540x_{*}^2-1-24x_{*}\sqrt{3(27x_{*}^2-1)^3}\bigg)^{1/3}\bigg].
\end{equation}
As this formula includes square root and cube root, we need to specify the argument of complex number to avoid the branch cut. Here we choose $\arg[3(27x_{*}^2-1)^3]=\{0,\pi\}$, $\arg[5832x_{*}^4+540x_{*}^2-1+24x_{*}\sqrt{3(27x_{*}^2-1)^3}]\in (-2\pi,0]$, and $\arg[5832x_{*}^4+540x_{*}^2-1-24x_{*}\sqrt{3(27x_{*}^2-1)^3}]\in [0,2\pi)$. And we can conclude that this phase boundary exists in the regime $x<\sqrt{3}/9$ and $y>8\sqrt{3}/9$. Therefore, only when $N\chi/\Omega>8\sqrt{3}/9$, the DPT occurs. Instead, if $N\chi/\Omega<8\sqrt{3}/9$, there is a smooth crossover. 

As we mentioned in the main text, our experiment always lie in the DPT regime, and we characterize the ferromagnetic and paramagentic phase using the long-time average of excitation fraction $\overline{N_{\uparrow}}/N$, which is possible to express in terms of $s^z$,
\begin{equation}
    \frac{\overline{N_{\uparrow}}}{N}=\frac{1}{2}(\overline{s^z}+1), \quad \overline{s^z}=\frac{1}{T}\int_0^Ts^z(t)\mathrm{d}t,
\end{equation}
where $T$ is the oscillation period of $s^z$. This integral can be evaluated using  Eq.(\ref{eq:dpt4}), as
\begin{equation}
    \int_0^Ts^z(t)\mathrm{d}t=\int_{-1}^{s^z_{*}}\frac{2s^z\mathrm{d}s^z}{\sqrt{-2V(s^z)}}, \quad T=\int_{-1}^{s^z_{*}}\frac{2\,\mathrm{d}s^z}{\sqrt{-2V(s^z)}}.
\end{equation}
All these integrals can be calculated analytically in terms of elliptic integrals. Considering the asymptotic behavior near the critical point, we find that instead of the sudden jump behavior of $s^z_{*}$, the long-time average $\overline{s^z}$ is continuous at the critical point, and the first derivative of $\overline{s^z}$ diverge logarithmically. The following formula describe the asymptotic behavior of $\overline{s^z}$ with fixed $N\chi/\Omega$,
\begin{equation}
    \overline{s^z}\rightarrow \overline{s^z_c}+\frac{C}{\ln|x-x_{*}|}, \quad \overline{s^z_c}=\frac{1}{2}-\frac{1}{2}\sqrt{1-\frac{8x_{*}}{y_{*}}},
\end{equation}
where $C$ is a constant set by $N\chi/\Omega$. We can verify  the asymptotic behavior predicted above numerically. This is plotted in Fig.~\ref{sfig2}(c) for the case $N\chi/\Omega=5$. The continuous behavior of $s^z_{*}$ at the critical point determines a second-order dynamical phase transition in our case.

Here we also discuss how to determine the critical point in experiment. Based on Fig.~\ref{sfig2}(c), the first derivative of the long-time average excitation fraction diverges at the critical point.
However, in the analysis of experimental data, we have to use the finite difference as an approximation of the first derivative, which is limited by the precision of laser frequency and experimental fluctuation. To construct a stable phase boundary, instead we use the maximum transfer point as a signature of the critical point. 
Also, as we mentioned in the main text, we measure the Rabi spectrum at a fixed time instead of taking the long-time average excitation fraction due to technical challenges. All these systematic errors in determination of the critical point are smaller than the measurement error bars under current experimental conditions.

Finally we discuss the scaling factor in $\chi$ used in the main text to match the experimental critical points to the DPT in LMG model. This scaling factor originates from the inhomogeneities in the Ising coulings $\chi_{ij}$, which couple the Dicke manifold to the states with different total spins. In this way, the effective Ising coupling should be modified by an overall factor from its value $\chi$ in Dicke manifold. We use the same scaling factor $0.56$ for all the measurements with different atom densities, and the experimental critical points agrees very well with the phase boundary in LMG model.

\section{Discussion of spin squeezing}
\begin{figure}[t]
    \includegraphics{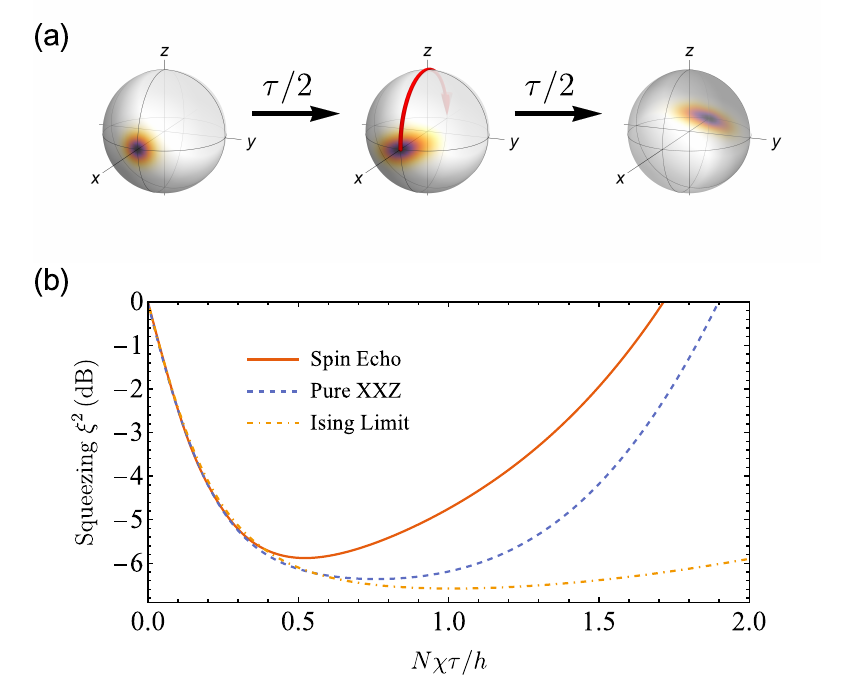}
    \caption{\label{sfig3}(a) Spin echo sequence for generation of squeezing using motional sidebands (see text). The total spin state $|\Psi(t)\rangle$ is illustrated using the Husimi-Q function \cite{ma2011}. (b) Comparison of the obtained spin squeezing for a spin echo sequence (DTWA method), the pure XXZ model (DTWA method) and the Ising limit (analytic solution). The spin squeezing parameter is expressed in terms of decibels (dB), i.e. $10\log_{10}\xi^2$.}
\end{figure}

We proceed to study the role of quantum correlations and entanglement in our XXZ simulator by theoretical calculation of spin squeezing, since it provides a relevant entanglement witness and an important resource for quantum metrology \cite{ma2011}. 
We study the proposed Ramsey spectroscopy sequence depicted in Fig.~\ref{sfig3}(a).
Initially all atoms are assumed to be in the $|\downarrow\rangle$ state and a $\pi/2$ blue-sideband pulse is applied to transfer the atoms to the $|S^x=N/2\rangle$ state. 
Then the system is allowed to evolve for $\tau/2$ under XXZ interaction (see Eq.(2) in the main text), followed by a blue-sideband spin echo pulse, and a further evolution time $\tau/2$.
The additional spin echo pulse at half of the evolution suppresses the dephasing effect of inhomogeneous longitudinal fields.
The squeezing is quantified by the Ramsey spin squeezing parameter \cite{wineland1992}, $\xi^2=\min\limits_{\theta}N(\Delta S_{\theta}^{\perp})^2/|\langle\mathbf{S}\rangle|^2$, which signals entanglement if $\xi^2<1$.
Here, $(\Delta S_{\theta}^{\perp})^2$ is the variance of the spin noise along an axis perpendicular to the collective spin $\langle\mathbf{S}\rangle$, parametrized by an angle $\theta\in[0,2\pi)$. This squeezing parameter can be extracted by an appropriate sequence of spin rotations at the end of the Ramsey protocol.

To estimate the achievable spin squeezing, we adopt the discrete truncated Wigner approximation (DTWA), which solves the mean-field equations of motion supplemented by Monte Carlo sampling of the initial conditions to account for quantum fluctuations \cite{schachenmayer2015}. 
We choose $N_{\mathrm{th}}=1000$ and scale the longitudinal field from the one in the experiment by a factor $N_{\mathrm{th}}/N_{\mathrm{exp}}$. The theoretical prediction of spin squeezing is depicted in Fig.~\ref{sfig3}(b). We compare it to the spin squeezing in the pure XXZ model (ignoring longitudinal fields in Eq.(2) in the main text), and the Ising model ($H_{\mathrm{Ising}}=\sum_{ij}\chi_{ij}S_i^zS_j^z$) which allows for an exact solution (see below). 
We find that the beyond-mean-field dynamics in our simulator is similar to the Ising limit, with an additional small suppression arising from the inhomogeneties in the longitudinal fields.
For $N_{\mathrm{th}}=1000$ atoms, near $6$dB optimal squeezing can be achieved at $N\chi\tau/h\approx 0.5$, which translates under current experimental conditions to optimal squeezing times around $100$ms.
On this time scale we do not expect detrimental effects from the technical imperfections.
The predicted squeezing emphasizes the metrological potential of motional sidebands.

\subsection{Analytic solution for Ising model}
In Fig.~\ref{sfig3}, we compare the theoretical calculation of the achievable spin squeezing in our XXZ simulator with the analytic solution of Ising model ($\hbar=1$),
\begin{equation}
H_{\mathrm{Ising}}=\sum_{ij}\chi_{ij}S_i^zS_j^z.
\end{equation}
Without loss of generality, we assume the Ising couplings are symmetric, $\chi_{ij}=\chi_{ji}$. As we start from the initial state $|S^x=N/2\rangle$, it is easy to show that the collective spin always stays in $x$ direction, and we can simplify the definition of spin squeezing parameter $\xi^2$ in the main text as follow,
\begin{equation}
\xi^2=\min_{\theta}\frac{N(\Delta S^{\perp}_{\theta})^2}{|\langle S^x\rangle|^2},
\end{equation}
where
\begin{equation}
(\Delta S^{\perp}_{\theta})^2=\cos^2\theta\langle S^zS^z\rangle+\sin^2\theta\langle S^yS^y\rangle+\cos\theta\sin\theta\langle S^zS^y+S^yS^z\rangle.
\end{equation}

Following the technique of discussing one-axis-twisting model in Ref.~\cite{ma2011}, all the expectation values above can be evaluated as follows,
\begin{equation}
    \begin{gathered}
    \langle S^x\rangle=\frac{1}{2}\sum_k\prod_{j}^{(k)}\cos(\chi_{kj}t), \quad\langle S^zS^z\rangle=\frac{N}{4},\\
    \langle S^yS^y\rangle=\frac{N}{4}+\frac{1}{4}\sum_{k<l}\bigg[\prod_j^{(k,l)}\cos\big[(\chi_{kj}-\chi_{lj})t\big]-\prod_j^{(k,l)}\cos\big[(\chi_{kj}+\chi_{lj})t\big]\bigg],\\
    \langle S^zS^y+S^yS^z\rangle=\frac{1}{2}\sum_{k<l}\sin(\chi_{kl}t)\bigg[\prod_j^{(k,l)}\cos(\chi_{kj}t)+\prod_j^{(k,l)}\cos(\chi_{lj}t)\bigg],\\
    \end{gathered}
\end{equation}
where $\prod_j^{(k)}$ means multiplication without the term $j=k$. Based on all these expectation values, we can calculate the spin squeezing parameter $\xi^2$ by tuning $\theta$ to reach the minimum value of $(\Delta S^{\perp}_{\theta})^2$. We define this angle as optimal squeezed angle $\theta_0$, and we get
\begin{equation}
    \begin{gathered}
    \min_{\theta}(\Delta S^{\perp}_{\theta})^2=\frac{A+B}{2}-\frac{1}{2}\sqrt{(B-A)^2+C^2}\\
    \tan(2\theta_0)=\frac{-C}{B-A}\\
    \end{gathered}
    \label{eq:min}
\end{equation}
where $A=\langle S^zS^z\rangle$, $B=\langle S^yS^y\rangle$, $C=\langle S^zS^y+S^yS^z\rangle$. In this way, the spin squeezing for general case of Ising interaction can be evaluated analytically.

\begin{figure}[t]
    \includegraphics{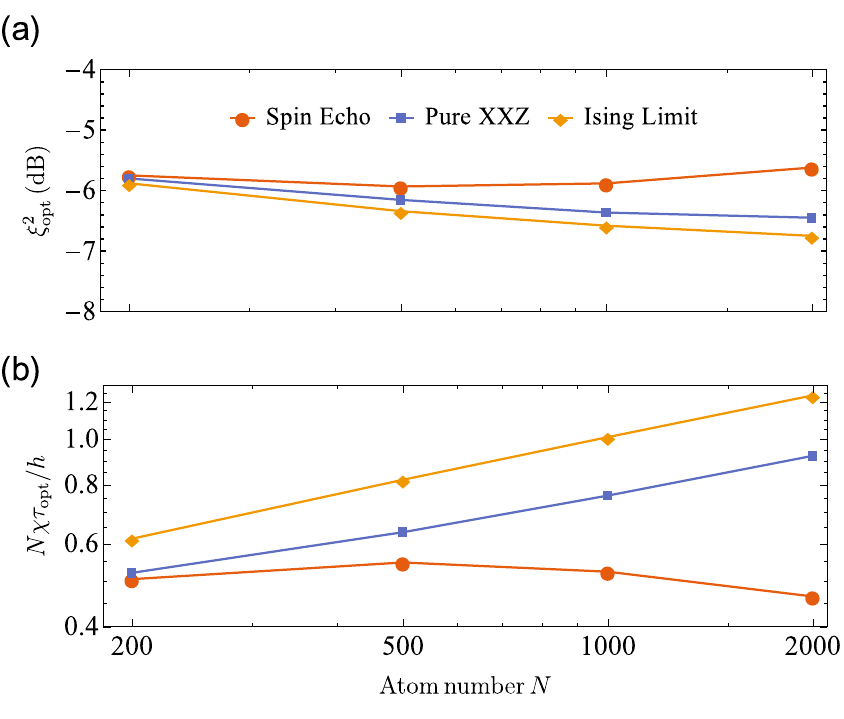}
    \caption{\label{sfig4}(a) Finite-size scaling of optimal spin squeezing in our XXZ simulator with spin echo sequence (DTWA method), compared to pure XXZ model (DTWA method) and Ising limit (analytic solution). (b) Finite-size scaling of optimal squeezed time in our XXZ simulator with spin echo sequence (DTWA method), compared to pure XXZ model (DTWA method) and Ising limit (analytic solution).}
\end{figure}

\subsection{Prediction of achievable spin squeezing}
In Fig.~\ref{sfig3}, we discuss the achievable spin squeezing in our XXZ simulator, with comparison to the pure XXZ model and the Ising limit. As it is hard to calculate spin squeezing for $\sim 10^5$ atoms in theory, we choose $N_{\mathrm{th}}=1000$ and scale the longitudinal field from the one in the experiment by a factor $N_{\mathrm{th}}/N_{\mathrm{exp}}$. The spin squeezing as a function of Ramsey dark time is depicted in the main text (see Fig.~4(b)). Here we use finite-size scaling as a way to predict the achievable spin squeezing under experimental conditions (see Fig.~\ref{sfig3}). 

We extract the optimal spin squeezing (see Fig.~\ref{sfig3}(a)) and optimal squeezed time (see Fig.~\ref{sfig3}(b)) with $N_{\mathrm{th}}=200, 500, 1000, 2000$. We find that under current experimental conditions, the optimal squeezing saturates around $6$dB when we increase the atom number in theory, and the optimal squeezed time stays near $N\chi\tau/h\approx 0.5$. Unfortunately the finite-size scaling curve for the current experiment condition is not monotonic, which means that our estimation of the optimal spin squeezing is not necessarily accurate. The analysis nevertheless shows the detrimental effects caused by the inhomogeneities in longitudinal fields, which will lead to a non-negligible suppression of spin squeezing, compared to pure XXZ model and Ising limit. Therefore we predict that if it were possible to carefully control the longitudinal fields in experiment and reduce their size, one could get closer to the finite-size scaling curve of pure XXZ model, which increases monotonically when increasing atom number, although the improvement is not significant due to the thermal distribution. In this case, we predict an optimal spin squeezing set by $\xi^2_{\mathrm{opt}}\propto N^{-0.067}$, and an optimal squeezed time by $\tau_{\mathrm{opt}}\propto N^{-0.752}$. Ideally speaking, for $5\times 10^5$ atoms, 8dB optimal squeezing can be achieved around $100$ms.